# Fiber Bragg Grating sensors for deformation monitoring of GEM foils in HEP detectors

L. Benussi[2], S. Bianco[2], M. Caponero[4], S. Muhammad[1,2,34], L. Passamonti[2]. D. Piccolo[2], D. Pierluigi[2], G. Raffone[2], A. Russo[2], G. Saviano[1,2]

[1] University of Rome "La Sapienza" (IT) - Facoltà di Ingegneria, Ingegneria Chimica Materiali ed Ambiente
[2] INFN-Laboratori Nazionali di Frascati Via E. Fermi 40, I-00044 Frascati, Rome, Italy
[3] National Centre for Physics, Quaid-i-Azam University Campus, Islamabad, Pakistan.

[4] Centro Ricerche ENEA Frascati, via E. Fermi 45, 00044 Frascati, Rome, Italy

*Abstract*—Fiber Bragg Grating (FBG) sensors have been so far mainly used in high energy physics (HEP) as high precision positioning and re-positioning sensors and as low cost, easy to mount, radiation hard and low space-consuming temperature and humidity devices. FBGs are also commonly used for very precise strain measurements. In this work we present a novel use of FBGs as flatness and mechanical tensioning sensors applied to the wide Gas Electron Multiplier (GEM) foils of the GE1/1 chambers of the Compact Muon Solenoid (CMS) experiment at Large Hadron Collider (LHC) of CERN. A network of FBG sensors has been used to determine the optimal mechanical tension applied and to characterize the mechanical stress applied to the foils. The preliminary results of the test performed on a full size GE1/1 final prototype and possible future developments will be discussed.

I. INTRODUCTION

The GE1/1 CMS upgrade consists of 144 GEM chambers of about 0.5 m² active area each and based on the triple GEMs technology, to be installed in the very forward region with pseudorapidity ($1.6 < \eta < 2.4$) of the CMS endcap [1]. The large active area of each GE1/1 chamber consists of three GEM foils (the largest GEM foils assembled and operated so far) to be mechanically tensioned in order to secure their flatness and the consequent uniform performance of the GE1/1 chamber across its whole active surface. To assure the design performance of the GEM detector, the uniform distance (1 mm) between the three GEM foils must be kept and monitored with a 30 μm accuracy).

FBG sensors are produced on standard telecommunication single mode optical fiber with acrylate coating. To make the sensor, the acrylate coating is removed at the selected location acrylate being usually adopted for temperature sensing, polyimide being usually adopted for structural sensing [2][3]. As FBGs are commonly used for very precise strain measurements. So in this work we present a novel use of FBGs as flatness and mechanical tensioning sensors and, as such, used to monitor the planarity of foils An FBG sensor is an optical fiber whose refractive index of the core varies periodically. This periodicity is normally obtained by irradiating a UV sensitive fiber with a spatially modulated intense UV laser beam [4]. When broad spectrum light is injected into fiber and interacts with the Braggs grating the wavelength is reflected in a narrow band ~0.1 nm. The Bragg wavelength $\lambda$

$$\lambda = 2n\Lambda \qquad (1)$$

which reflects from grating is a function of effect refractive index n and grating pitch $\Lambda$. Thus, we can measure the strain which deforms the grating pitch by measuring the Bragg wavelength [5].

FBG sensors have been installed on the three GEM foils at different positions in transverse and longitudinal direction to monitor the deformation when foils are stretched or loosen simultaneously. The effect of temperature variations on FBG sensors is canceled out by normalization with FBG sensors not mechanically attached to the detector foils but outside.

To monitor the vertical displacement of the top foil during the test we are using a Laser Displacement Sensor (LDS). Due to the localization limit of the LDS, in parallel a Moiré fringes pattern setup is built which is used to measure the overall deformation in the top GEM foil. These three techniques are being applied simultaneously to cross check each other and to make sure the uniformity and flatness of the GEM foils.

## II. FBG Sensors setup for deformation measurement of GE1/1 foils

The FBG sensors are glued on the three GE1/1 films in transverse and longitudinal directions. Glues tested include UHU PWS 24h, 2011 ARALDITE HUNTSMAN, PATTEX PLASTIC HENKEL, and UV-RAY WELLOMER UV4028. Glue selected was 2011 ARALDITE whose mechanical properties and radiation hardness are well known. A suitable set of tools and procedures was developed to assure reliable mechanical strength, while still retaining the requirement of minimal glue deposition.

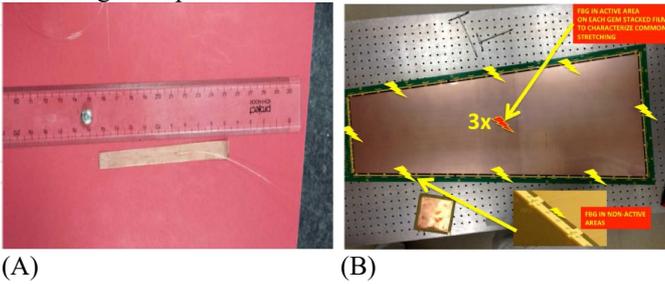

(A)  (B)

Figure 1. (A). gluing of a FBG sensor on GEM sample.
(B). FBG sensors on GEM foils in a GE1/1 chamber

The sensors in the middle of GEM planes are used once to certify the uniformity of stretching procedure over the three GEM planes. Sensors installed on the upper GEM plane only, provide inter calibration with Moiré's and LDS systems, and deformation monitoring.

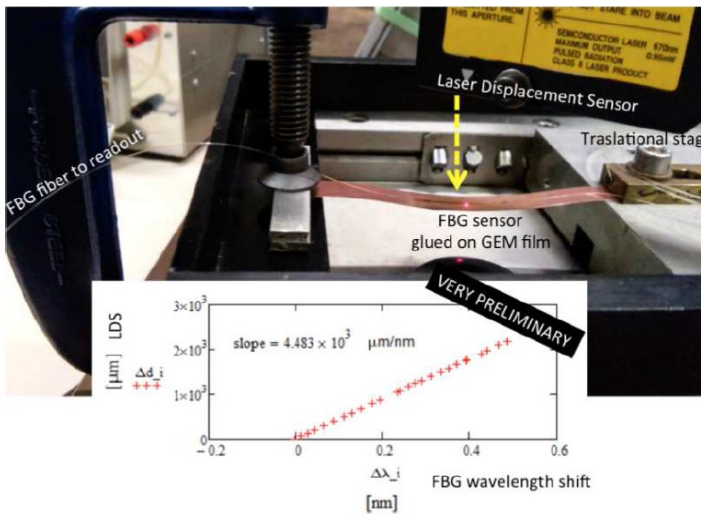

Figure 2. Test of gluing a FBG sensor on a GEM film strip. The FBG response is very well correlated with the gravitational Sagitta as measured by LDS.

Illustration shows the experimental setup with LDS (top), translational stage pulling the GEM foil strip (right), FBG sensor glued on GEM foil strip (centre) and optical fibre funnelling the laser light to interrogation system (left).

## III. Moirè fringes pattern method to study the flateness of the top gem film

The developments in fringe processing have provided an automated and accurate way to analyse the fringe patterns resulting from optical methods for displacement measurement [6]. Although a variety of methods for processing fringe patterns exists, there are two fundamental methods that are predominantly used. One of these methods consists of extracting skeleton lines from the fringe patterns to enable interpolation of the whole-field displacement. The second method is based on phase shifting. Through the use of a phase shifter [e.g., a piezoelectric transducer (PZT)], a known uniform phase is added to the interferometric fringe patterns to obtain a series of fringe pattern images under the same magnitude of applied load. Using specified formulas, the displacement distribution can then be determined from pixel grayscales [7].

The projector (mounted on a translational stage for phase-shift algorithm) illuminates the GE1/1 with a pattern generated via a Ronchi grating. The receiver lens is equipped with the same Ronchi grating. Moiré fringes are generated on the lens focal plane proportional to the GE1/1 non-planarity.

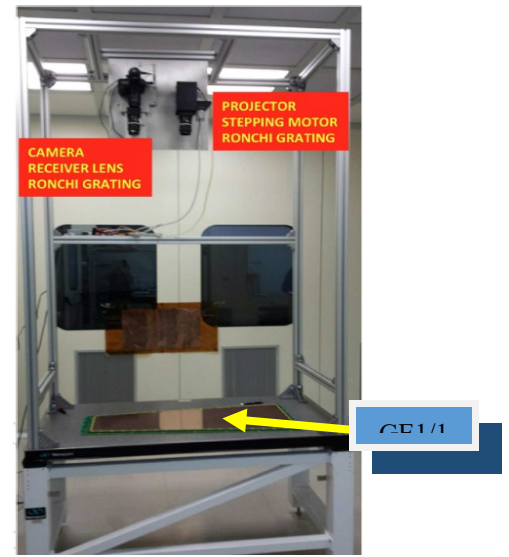

Figure 3. Moiré setup in Frascati clean room projecting fringes on a whole GE1/1 detector.

Shadow Moiré is a method used to map out of plan contours of a surface [8, 9]. This method involves placing a grating of pitch P in front of the surface to be contoured and the illuminating it with collimated light. When this grating and the shadow of this grating are viewed together an interference pattern is produced that is out of plan contour.

## Iς. Preliminary Results and Discussion

During the test, the mechanical tension of the GEM foil is varied over time from a non-tensioned state to a tensioned state. Two sets of sensors are installed, i.e. transverse and longitudinal to the GEM foil. Each set is composed of three sensors each glued on a GEM foil. The sensors output (shown in strain units) is very consistent and uniform during the foil stretching (Figure4),o and in the final stretched state, the srain variation is equal to 0.05mstrain which is good. This preliminary result is a strong and solid indication that the three GEM foils are subject to the same tensile load during the defined assembly procedure.

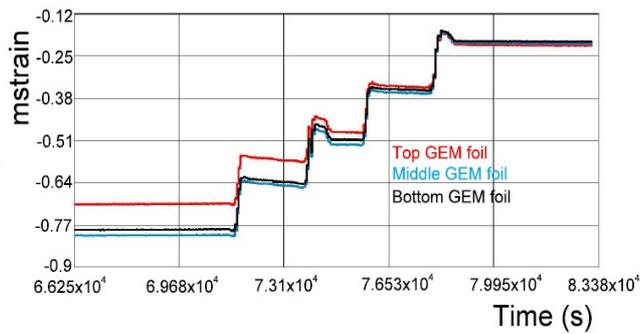

Figure 4. Preliminary data on the FBG sensors output during a tensioning cycle.

From this test our foils stretching technique is validated, as tension increased the foils align themselves and behave uniformly. Instead in loose condition the foils are curved like concave or convex and strain variation is high as can be seen in start of the above plot.

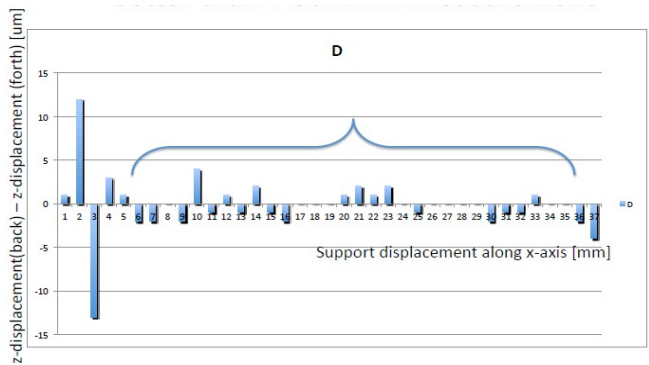

Figure 5. Repeatability of LDS system over 37 mm scan is better than 4-5µm in Z measurement.

Residuals for a back-forth scanning of reference surface with the Laser Displacement System used to calibrate the Moiré's fringes. Repeatability of LDS system over a 37mm scan is better than about 4 $\mu m$ in the measurement of z direction (transversal to scan) displacement as shown in Figure 5.

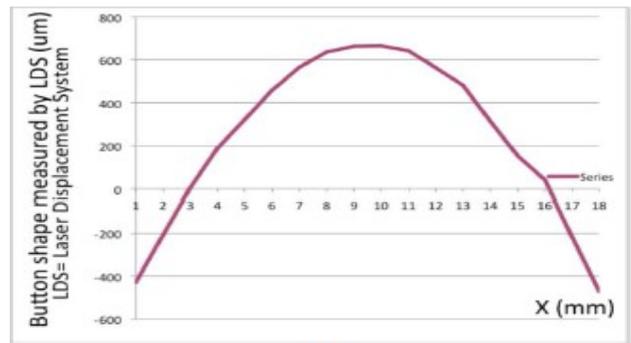

Figure 6. Button shape measurment by LDS

As mentioned above we are using three systems parallel in order to make sure the uniformity of the GE1/1 foils, figure 6 and 7 are the LDS and Moire fringes test for the known surface, it is very important to mutually calibrate the two setups, which will cross check the FBG sensors results only for the top GEM foil.

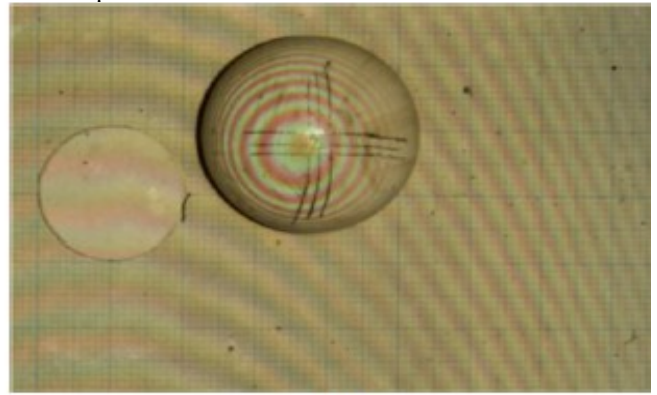

Figure 7. Fringes on the button "Moire fringes"

One-fourth of period is easily visible, hence the estimate on resolution is 100 $\mu m$. Finer grating and phase shift algorithm will improve resolution to better than 30 $\mu m$.

ς. CONCLUSIONS

ςI.

We proved that use of a distributed FBG sensor network can effectively address the definition of best procedure to be adopted for the correct stretching of the three GEM foils while assembling the GE1/1 detector. LDS was used to measure punctual Sagitta in order to both validate the technique adopted for FBG sensor installation and to perform local control of FBG operation during experimental tests.

Use of Moiré fringes was considered for non contact control of GEM foil planarity, aiming to the developed of a setup to be used on-line in production of the GE1/1 detector. Work is in progress for the development of such setup, with custom engineered designed and phase-shift algorithm capability to improve resolution.

Results show that by the developed assembling procedure the three GEM foils are stretched simultaneously and to a similar final state, even if previously laying in the frame in unevenly loose condition. Use of FBG sensors allowed to verify what stress is sufficient to make GEM foils be plane stretched, which is important not to prescribe overestimated tensioning that would uselessly cause overstress and risk of mechanical damage of GEM foils.
In addition as described above very nice